\documentclass[journal]{IEEEtran}

\usepackage[scriptsize]{caption}
\usepackage{graphicx, dblfloatfix}
\usepackage{epstopdf}
\usepackage{amsmath}
\usepackage{amssymb}
\usepackage{graphics}
\usepackage{cite}
\usepackage{amsfonts}
\usepackage{textcomp}
\usepackage{multicol}
\usepackage{multirow}
\usepackage{wrapfig}
\usepackage{color}
\usepackage{fixltx2e}
\usepackage{subfig}
\ifCLASSINFOpdf
\else
\fi
\begin{document}

\title{Demonstration of an LO-less, DSP-free QPSK Receiver for Data Center Interconnects}

\author{Rashmi Kamran\textsuperscript{(1)}, Nandakumar Nambath\textsuperscript{(2)}, Sarath Manikandan\textsuperscript{(1)}, Rakesh Ashok\textsuperscript{(1)}, Nandish Bharat Thaker\textsuperscript{(1)}, Mehul Anghan\textsuperscript{(1)}, and Shalabh Gupta\textsuperscript{(1)}\\\textsuperscript{(1)}\textit{Department of Electrical Engineering, IIT Bombay, Mumbai-400076, India}; \textsuperscript{(2)}\textit{School of Electrical Sciences, IIT Goa, Ponda-403401, India}}
\vspace{-0.5cm}
\maketitle
\begin{abstract} We present the first demonstration of a local oscillator (LO)-less digital signal processing (DSP)-free coherent receiver for high-capacity short distance optical links. Experimental results with an analog domain constant modulous algorithm (CMA)-based equalizer chip for the self-homodyne quadrature phase shift keying (SH-QPSK) system validate the employability of an all-analog and LO-less receiver for low-power interconnects.
\end{abstract}
\IEEEpeerreviewmaketitle

\section{Introduction}
The Ethernet roadmap targets interface speeds of up-to 1.6\,Tb/s for future data center interconnects, in which reduction of power dissipation is a major design consideration \cite{ethernet}. Usage of optical self-homodyne (SH) techniques is an attractive solution for reducing power consumption of the receiver electronics for data center interconnects at the targeted interface speeds. SH systems employ local oscillator (LO)-less coherent receiver and require simpler signal processing as opposed to the commonly used power hungry digital signal processing (DSP) based coherent receivers. SH systems do not have stringent linewidth requirements, and hence do not need an expensive laser at the transmitter \cite{WHYSH6}. SH systems also facilitate the deployment of spectrally efficient techniques for high capacity links \cite{16QAM1,MPSK}. Due to the reduced signal processing requirements, optical channel impairment compensation can easily be performed by an analog domain equalizer in such systems.  
\par DSP based coherent receivers consume excessive amounts of power as they require high-speed analog-to-digital converters (ADCs) and digital signal processors \cite{CPD}. An analog-domain signal processor (ASP) obviates the need for power hungry ADCs and DSP. For example, an ASP based receiver in a 130\,nm BiCMOS technology would require approximately 3.5\,W for 100\,Gb/s\,dual polarization-quadrature phase shift keying (DP-QPSK) system \cite{ADP2} for short distance links. Power consumption can be reduced further in advanced CMOS technologies. Hence ASP based receivers provide substantial power savings over the DSP based receivers.  ASP based solutions can also be used for higher order modulation formats, such as 16-QAM, to achieve high capacity short distance interconnects \cite{ASPQAM}.
 \par In this paper, for the first time, we demonstrate an LO-less QPSK receiver with polarization-multiplexed carrier and an analog-domain equalizer. A 20\,Gb/s SH-QPSK system with a CMA equalizer chip is experimentally demonstrated for optical links having up to 80\,km length of standard single mode fiber (SSMF).

\section{QPSK system with LO-less receiver}
 \begin{figure}[b!]
	\centering
	\includegraphics[width=0.95\columnwidth]{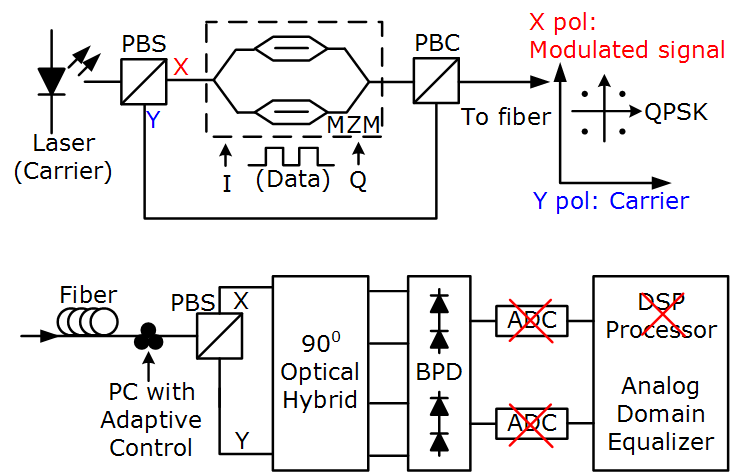}
	\caption{Architecture of the proposed SH-QPSK system with an analog domain equalizer and LO-less receiver. PBS/PBC: polarization beam splitter/combiner; MZM: Mach-Zehnder modulator; PC: polarization controller; BPD: balanced photo detector; and ADC: analog-to-digital converter. }
	\label{block}   
\end{figure} 

The architecture of the proposed system is illustrated in Fig.\,\ref{block}. It is a polarization diversity based SH system, which uses polarization multiplexed carrier for coherent detection. In the transmitter, the laser output is split into two orthogonal polarizations. The carrier in one of the polarizations is QPSK modulated using a nested Mach-Zehnder modulator. The modulated signal and the unmodulated carrier are polarization multiplexed using a polarization beam combiner (PBC) and sent through the fiber channel. 
\par At the receiver end, adaptive polarization control is required for separating the modulated signal and the carrier \cite{mehul1}. As polarization mode dispersion (PMD) is not significant for short-distance links, by maximizing received power in one of the polarizations (thereby minimizing it in the other one), the signal and the carrier can be separated out easily. The optical signal at the output of the polarization controller (PC) is given to a polarization beam splitter (PBS) to obtain the carrier (as the LO) and the modulated signal in two separate fibers. These outputs are fed to the optical hybrid followed by balanced photo-detectors (BPDs) for coherent detection. The electrical outputs from the BPDs are applied to an analog domain equalizer for compensating for the chromatic dispersion. 

\section{All analog domain (DSP-free) equalizer}
\begin{figure}[t!]
	\centering
	\includegraphics[width=0.75\columnwidth]{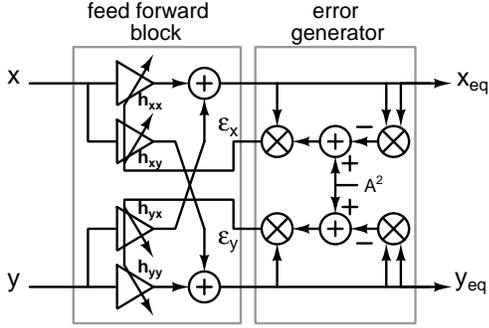}
	\caption{Architecture of the all-analog CMA equalizer\cite{firstofc2}. $x$ and $y$: unequalized input signals; $\mathbf{h}_{xx}, \mathbf{h}_{xy}, \mathbf{h}_{yx},$ and  $\mathbf{h}_{yy}$: filter coefficients; $\varepsilon_x$ and $\varepsilon_y$: error values; $A$: expected amplitude of the outputs; and $x_{eq}$ and $y_{eq}$: equalized output signals.}
	\label{chip}   
\end{figure}

The equalizer based on a blind CMA algorithm can be considered optimal for phase modulated signals. The analog-domain CMA equalizer which is designed for DP-QPSK systems \cite{firstofc2} is used with the SH-QPSK system. The equalizer has feed forward taps for X and Y polarization signals as well as cross taps between both the polarization signals, as presented in Fig.\,\ref{chip}, to mitigate the effects of PMD (although cross-taps are not really required in the proposed system because of external polarization control). The tap coefficients ($\mathbf{h}_{xx}, \mathbf{h}_{xy}, \mathbf{h}_{yx}$, and $\mathbf{h}_{yy}$) are adapted for minimizing errors ${\varepsilon_x=x_{eq}[A^2-|x_{eq}|^2]}$ and ${\varepsilon_y=y_{eq}[A^2-|y_{eq}|^2]}$. Here ${x_{eq}}$ and ${y_{eq}}$ are the complex equalized signals and $A^2$ is the square of the expected amplitude $A$.
\begin{figure}[t!]
	\centering
	\includegraphics[width=0.75\columnwidth]{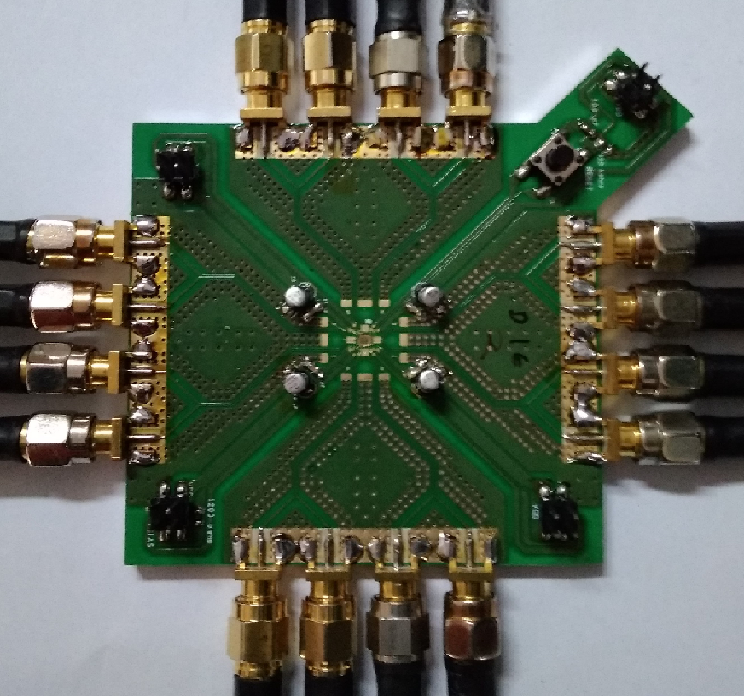}
	\caption{Printed circuit board with the CMA equalizer die bond-wired on it.}
	\label{board}   
\end{figure}
 \par This chip was designed in a 130\,nm BiCMOS technology from ST Microelectronics. It has two taps in each signal path with 20\,ps delay in between and was specifically designed for 100\,Gb/s DP-QPSK signals \cite{ADP2}. The chip currently dissipates 2.5\,W of power, which can be drastically reduced with implementation in advanced CMOS technologies. The high frequency printed circuit board (PCB) presented in Fig.\,\ref{board} is designed with Rogers RT/duroid 6010M laminate. All the on-board transmission lines have been designed as differential lines having a broadband frequency response from DC to 20\,GHz. The PCB contains decoupling capacitors as well as a reset switch to initialize the tap coefficient values. However, other assembly limitations such as impedance discontinuities due to bond-wire inductances limit the high-frequency operation. In the results presented in this paper the chip is tested for a 20\,Gb/s QPSK system with an LO-less receiver.
 
\section{Experimental setup and measurement results}
\begin{figure}[t!]
	\centering
	\includegraphics[width=\columnwidth]{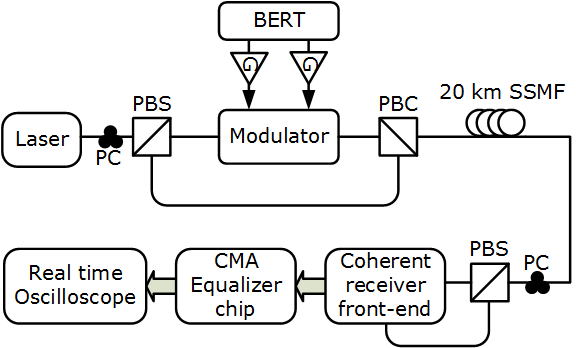}
	\caption{Experimental setup for a 20\,Gb/s SH-QPSK system with the CMA equalizer chip. PC: polarization controller; PBS/PBC: polarization beam splitter/combiner; BERT: bit error rate tester; SSMF: standard single mode fiber; and VOA: variable optical attenuator.}
	\label{SHQPSK}   
\end{figure} 
\begin{figure}[t!]
	\centering{
		\begin{tabular}{cc}
			\hspace{-0.4cm}\includegraphics[width=4.1cm,height=3.5cm]{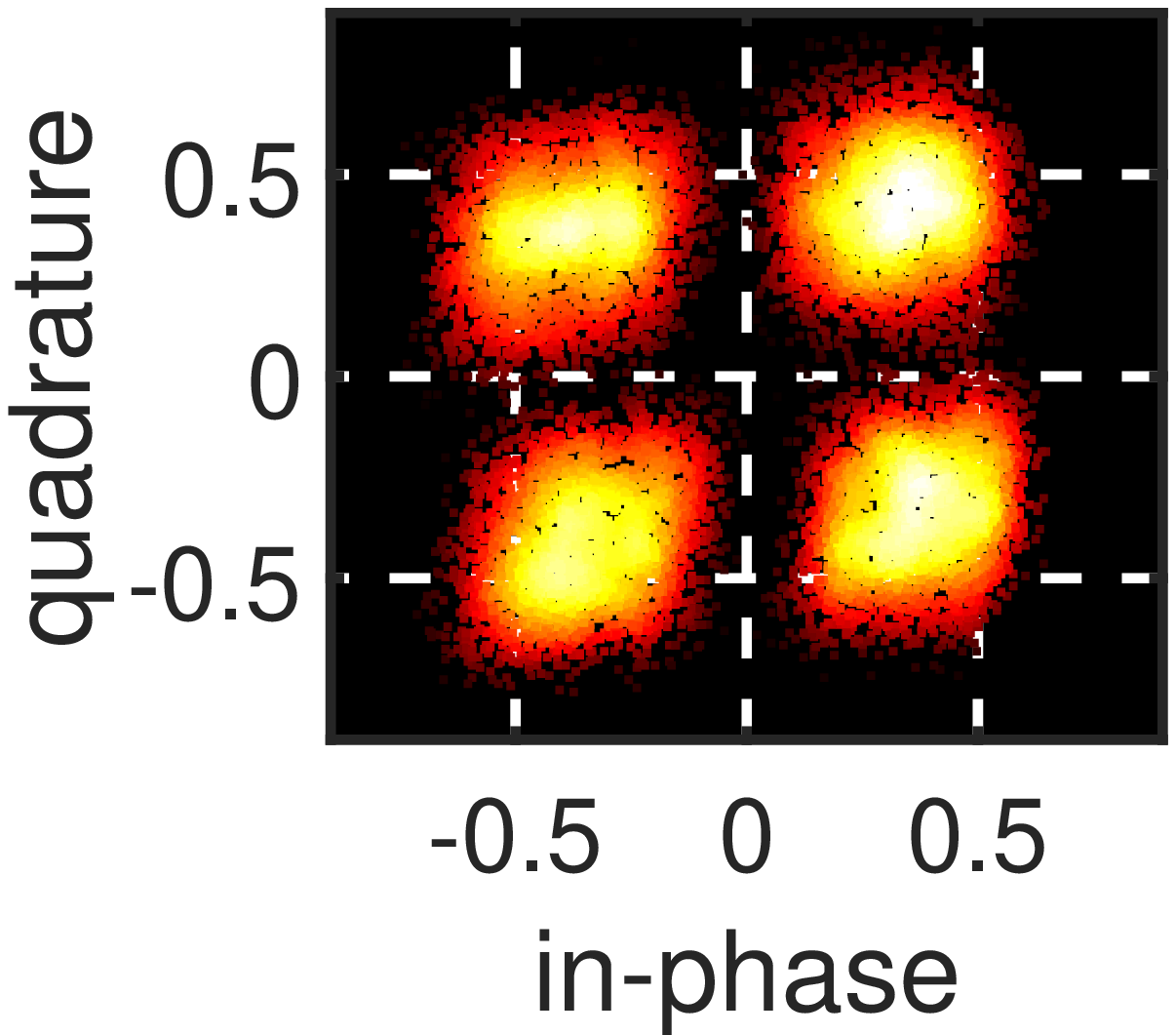}&
			\includegraphics[width=3.5cm,height=3.5cm]{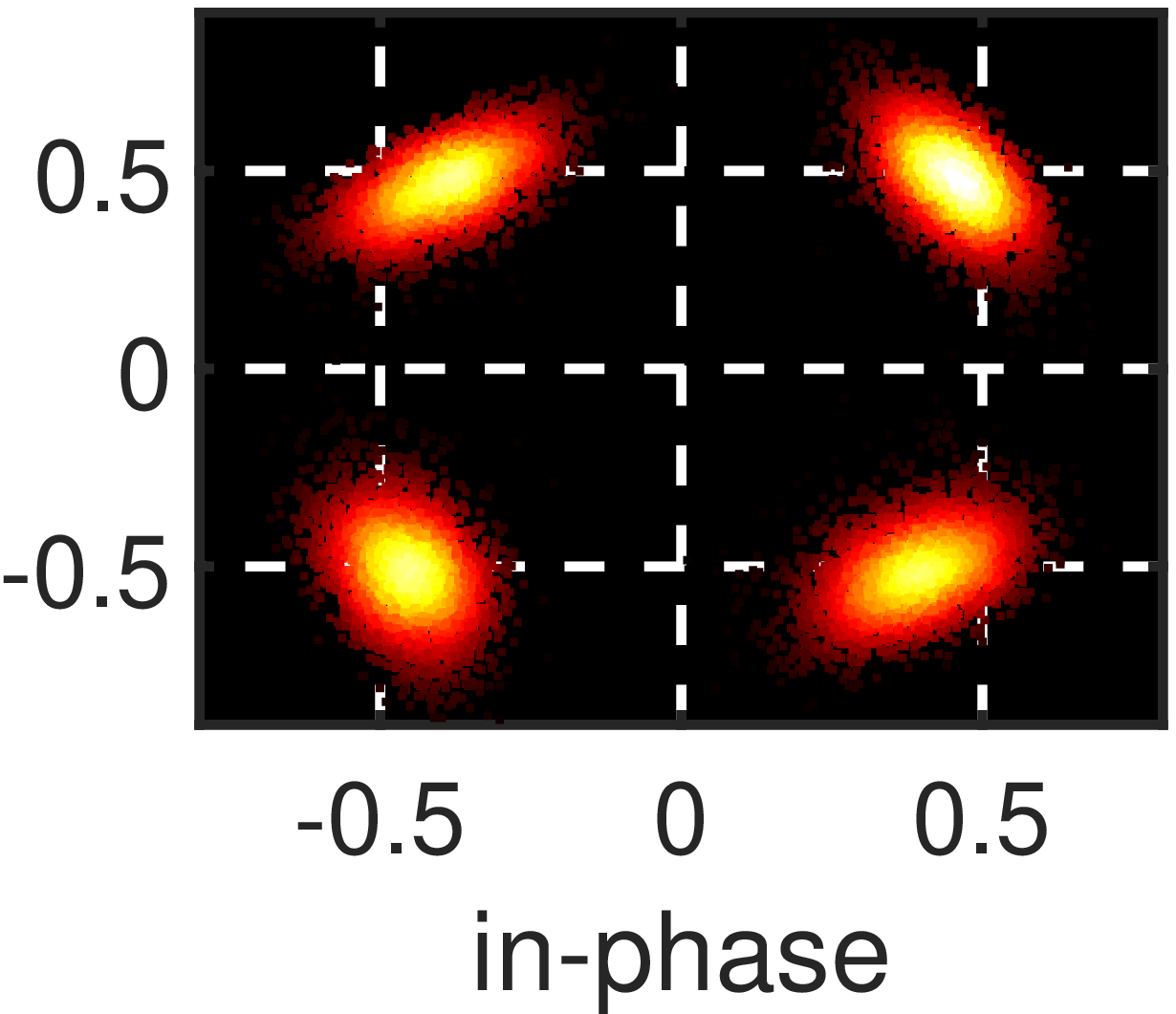}\\
			\hspace{-0.4cm}\scriptsize (a) &\hspace{-0.7cm} \scriptsize (b)\\[6pt]
			\hspace{-0.4cm}\includegraphics[width=4.1cm,height=3.5cm]{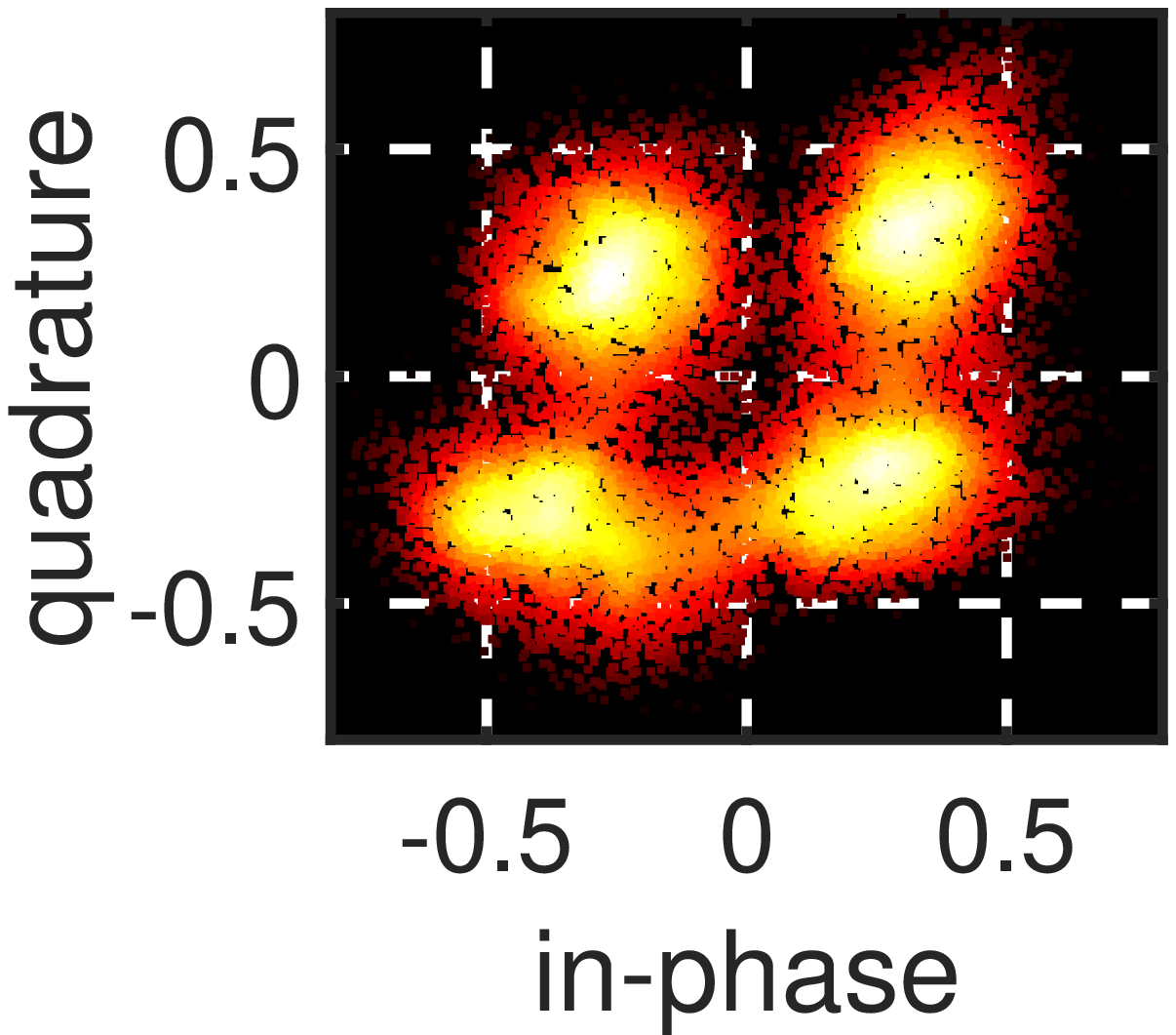}&
			\includegraphics[width=3.5cm,height=3.5cm]{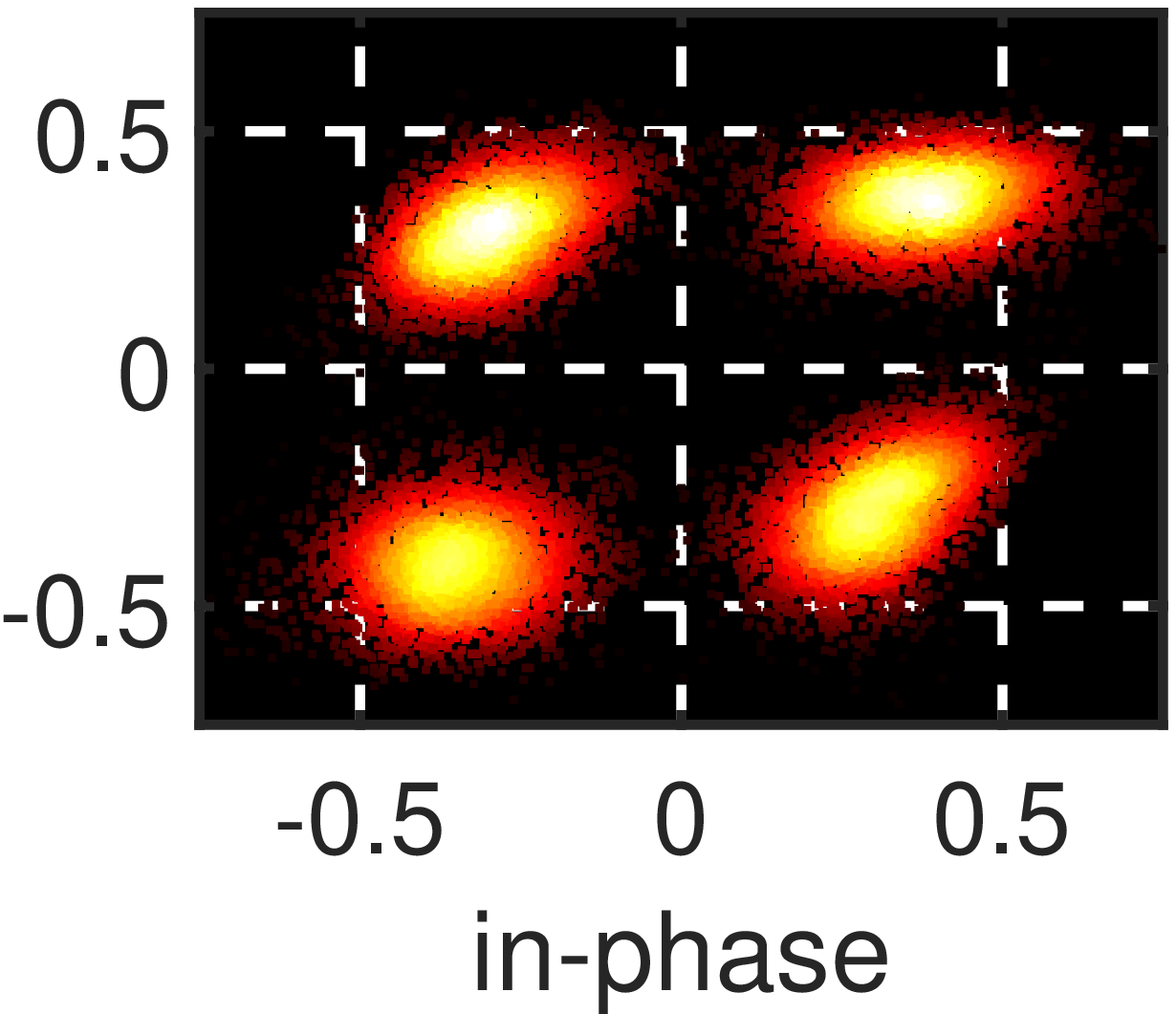}\\
			\hspace{-0.4cm}\scriptsize (c) &\hspace{-0.7cm} \scriptsize (d)\\[6pt]
			\hspace{-0.4cm}\includegraphics[width=4.1cm,height=3.5cm]{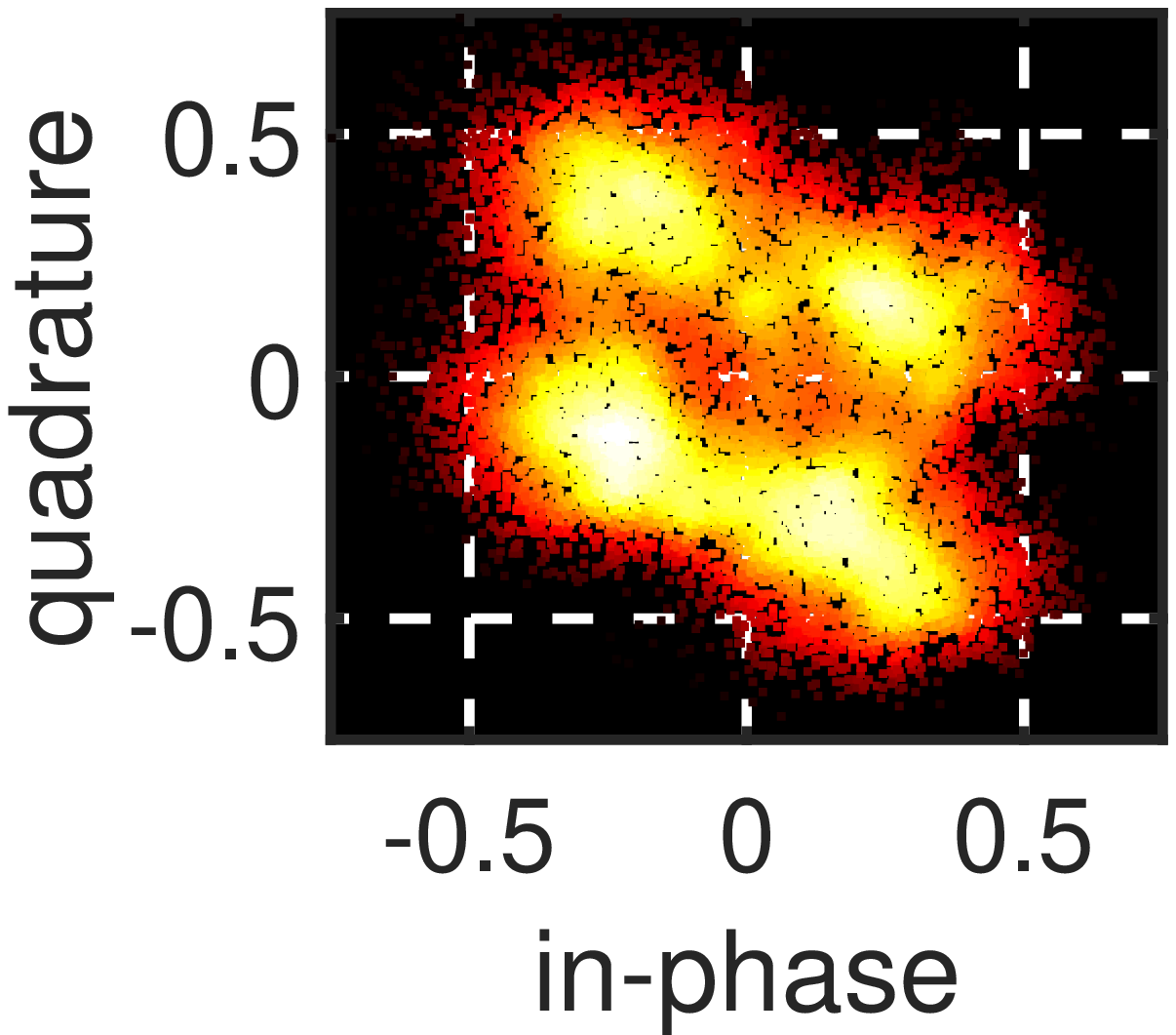}&
			\includegraphics[width=3.5cm,height=3.5cm]{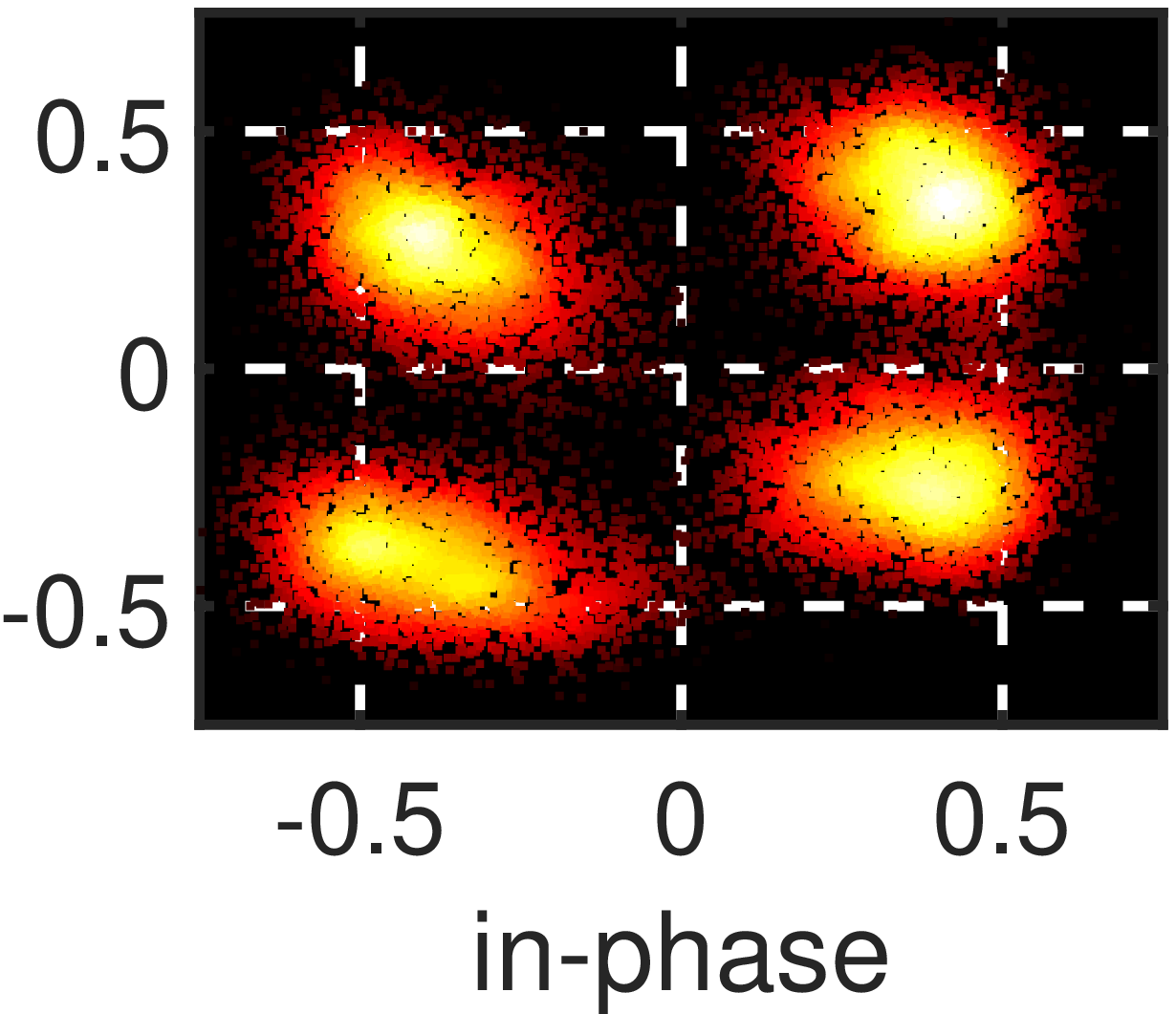}\\
			\hspace{-0.4cm}\scriptsize (e) & \hspace{-0.7cm}\scriptsize (f)
		\end{tabular}}
		\caption{Received constellations for a 20\,Gb/s SH-QPSK system with (a) Back-to-back link (BER = $ {1.01\times10^{-2}}$); (b) Back-to-back link with equalizer chip (BER = $ {2.7\times10^{-5}}$); (c) 20\,km link; (d) 20\,km link with equalizer chip (BER = $ {1.6\times10^{-3}}$); (e) 80\,km link; and (f) 80\,km link with equalizer chip (BER = $ {1.18\times10^{-2}}$).}
		\label{r1}
	\end{figure}

The experimental setup for the 20\,Gb/s QPSK system with the CMA equalizer chip is shown in Fig.\,\ref{SHQPSK}. An external cavity laser from Thorlabs (SFL1550P) is used for generating a 13.4\,dBm optical carrier at 1550\,nm wavelength. The carrier is split equally into two polarizations. One of the polarization components is applied to a QPSK modulator (LN86S-FC) and modulated using two streams of 10\,Gb/s data, generated by a bit error rate tester (N4962A), after amplification. The carrier from the other polarization has been multiplexed with the modulator output and sent to the fiber channel of different lengths. An optical amplifier is used for fiber length of 80\,km to compensate for the losses in the channel. At the receiver end, an external PC is used to manually control the output polarization to maximize power in one of the polarizations and minimize it in the other one (in a practical system, this can be achieved by using adaptive polarization control \cite{mehul1}). These polarizations are separated by a PBS and sent to the signal and LO ports of an integrated DP-QPSK coherent receiver front-end (CPRVx1222A) for the detection of SH-QPSK signals. By adjusting gain controls, single ended outputs of the receiver front-end are maintained at 400\,mVp-p to drive the CMA equalizer chip. A 33\,GHz digital signal analyzer from Keysight (OSAV334A) is used to capture the output signals from the chip. The optical modulator driver amplifiers (HMC788LP2E) provide reduced gain at high frequencies which results in poor quality of received signals, even for the back-to-back link experiment. This can be observed in the constellations shown in Fig. \ref{r1}a. The analog domain CMA equalizer improves the constellation significantly as shown in Fig. \ref{r1}b. Similarly, it can be observed in Fig. \ref{r1}c-f that the equalizer achieves a substantial improvement in the received constellations for fiber channel lengths of 20\,km and 80\,km.

Since the chip is designed for shorter distances and symbol periods (with only two taps), and due to the limitations in the experimental setup, limited bit error rates (BER) are achieved. Nevertheless, this demonstration qualitatively validates the usefulness of the analog-domain equalizer chip with an LO-less receiver.

\section{Conclusion}
In this paper, an LO-less SH-QPSK receiver with an all-analog (DSP-free) CMA equalizer chip has been presented.  A specifically designed analog domain equalizer for an SH receiver with polarization-multiplexed carrier can lead to achieve high-capacity low-power data center interconnects.

\section*{Acknowledgment}{The authors thank MeitY for supporting this work. We also thank Dr. Arvind Mishra and Dr. Madhan Thollabandi from Sterlite technologies, Aurangabad, for their valuable feedback.
}
\bibliographystyle{IEEEtran}
\bibliography{references4}

\end{document}